\documentclass{article}

\usepackage[english]{babel}

\usepackage[letterpaper,top=2cm,bottom=2cm,left=3cm,right=3cm,marginparwidth=1.75cm]{geometry}

\usepackage{amsmath}
\usepackage{graphicx}
\usepackage{caption}
\usepackage{graphicx, subfig}
\usepackage{upgreek}
\usepackage{tipa}
\usepackage{indentfirst}
\usepackage{setspace}
\usepackage{cite}
\usepackage[labelfont=bf]{caption}
\usepackage[colorlinks=true, allcolors=blue]{hyperref}
\usepackage{float}

\title{\textbf{Electronic properties of monolayer copper selenide with one-dimensional moiré patterns}}
\author{Gefei Niu$^{1\#}$, Jianchen Lu$^{1\#}$, Jianqun Geng$^{1\#}$, Shicheng Li$^1$, Hui Zhang$^1$, Wei Xiong$^1$, \\Zilin Ruan$^1$, Yong Zhang$^1$, Boyu Fu$^1$, Lei Gao$^{2\ast}$, Jinming Cai$^{1\ast}$}
\date{%
    $^1$Faculty of Materials Science and Engineering, Kunming University of Science and Technology, No. 68 Wenchang Road, Kunming 650093, China.\\%
    $^2$Faculty of Science, Kunming University of Science and Technology, No. 727 Jingming South Road, Kunming 650500, China.\\[2ex]%
    \leftline{$^{\#}$ These authors contributed equally.}
    \leftline{$^{\ast}$ Corresponding Authors: lgao@kust.edu.cn, jclu@kust.edu.cn, j.cai@kust.edu.cn}
}

\begin{document}
\maketitle

\begin{abstract}
Strain engineering is a vital way to manipulate the electronic properties of two-dimensional (2D) materials. As a typical representative of transition metal mono-chalcogenides (TMMs), a honeycomb CuSe monolayer features with one-dimensional (1D) moiré patterns owing to the uniaxial strain along one of three equivalent orientations of Cu(111) substrates. Here, by combining low-temperature scanning tunneling microscopy/spectroscopy (STM/S) experiments and density functional theory (DFT) calculations, we systematically investigate the electronic properties of the strained CuSe monolayer on the Cu(111) substrate. Our results show the semiconducting feature of CuSe monolayer with a band gap of 1.28 eV and the 1D periodical modulation of electronic properties by the 1D moiré patterns. Except for the uniaxially strained CuSe monolayer, we observed domain boundary and line defects in the CuSe monolayer, where the biaxial-strain and strain-free conditions can be investigated respectively. STS measurements for the three different strain regions show that the first peak in conduction band will move downward with the increasing strain. DFT calculations based on the three CuSe atomic models with different strain inside reproduced the peak movement. The present findings not only enrich the fundamental comprehension toward the influence of strain on electronic properties at 2D limit, but also offer the benchmark for the development of 2D semiconductor materials.
\end{abstract}


\section{Introduction}

Two-dimensional (2D) materials have drawn much attention in the past few years, owing to their novel physical and chemical properties and potential applications in various fields\cite{RN1,RN2,RN3,RN4,RN5,RN6,RN7}. As a member of the 2D material family, monolayer transition metal mono chalcogenides (TMMs) have sparked tremendous interest in recent years\cite{RN8,RN9,RN10,RN11,RN12,RN13,RN14,RN15,RN16,RN17}. Monolayer CuSe with graphene-like honeycomb lattice, a typical representative of 2D TMMs, had been theoretical predicted to possess 2D Dirac nodal line fermion (DNLF) protected by mirror reflection symmetry, making CuSe monolayer a new platform to study 2D DNLFs and potential applications in the field of high-speed low-dissipation devices\cite{RN12,RN16}. 

Moiré patterns formed by the lattice mismatch between the attached 2D materials and underlying substrates can introduce additional periodic potential field into 2D materials, which are expected to efficiently modulate the electronic properties of 2D materials\cite{RN18,RN19,RN20,RN21}. Normally, the symmetry of moiré patterns is highly related to the 2D materials and the substrate. For instance, the 3-fold symmetry of graphene grown epitaxially on the 6-fold symmetry of Ru(0001) substrate will produce the 6-fold symmetry of moiré patterns\cite{RN22}. However, CuSe monolayer grown on Cu(111) substrate presents a completely different situation. Due to the strong interaction between CuSe monolayer and Cu(111) substrate, the honeycomb lattice structure of CuSe monolayer tends to be stretched in one direction by a uniaxial strain, resulting in the formation of an unprecedented one-dimensional (1D) moiré patterns\cite{RN12}. It had been reported that 2D materials with 1D moiré patterns can host 1D topological interface mode arrays at smooth interfaces between local regions of distinct topological properties across the moiré\cite{RN23,RN24}. CuSe monolayer with 1D moiré patterns hosts a uniaxial strain. It is natural to ask how strain affects the electronic properties of CuSe monolayer on the Cu(111) substrate.

Here, we thoroughly investigated the effect of strain on the electronic properties of the CuSe monolayer with 1D moiré patterns on the Cu(111) substrate by a combination of scanning tunneling microscopy/spectroscopy (STM/S) and density functional theory (DFT). Our results show that three distinct regions of CuSe monolayer with 1D moiré patterns on the Cu(111) substrate have similar electronic properties, but a slightly different peak intensity at -0.66 eV, presenting a periodically modulated electronic properties of CuSe monolayer. We observe the domain boundary between two different orientations of 1D moiré patterns, and confirm that the boundary region suffers from a biaxial strain. STS measurements for the uniaxial strain, the biaxial strain and the strain-free regions (near the line defect) reveal the first peak in the conduction band will move slightly. The calculated local density of states (LDOS) for three models of CuSe monolayer with different strains reproduced the experimental STS spectra.

\section{Methods}

\subsection{Sample preparation and characterization}

Our experiment was performed by a commercial LT-STM/STS system (Scienta Omicron) operating at a base pressure below 1.0×10-10 mbar, equipped with a standard molecular beam epitaxy (MBE) system. All STM and STS results were acquired at liquid-helium temperature (4.2K). Cu(111) single crystal substrate was cleaned by cycles of argon-ion sputtering and subsequent annealing to 773 K. The Cu(111) surface quality was clarified through STM imaging. High purity Se powder (99.99\%, Sigma-Aldrich) evaporated from a Knudsen cell at ~393 K was deposited onto the room-temperature Cu(111) substrate. Subsequently, the sample was annealed to 473 K for 20 min to grow CuSe monolayer with 1D moiré patterns by a one-step selenization of Cu(111) substrate. All STM images were acquired in a constant-current mode, using an electrochemically etched tungsten tip. All bias voltages were applied to the sample with respect to the tip. The Nanotec Electronica WSxM software was used to process the STM images\cite{RN25}. dI/dV spectra were recorded by a lock-in amplifier with a bias modulation of 10 mV and a frequency of 599 Hz.

\subsection{First-principle calculations}

First-principles calculations were performed by using the Vienna ab initio simulation package (VASP). The projector augmented wave method was employed to describe the core electrons. The local density approximation (LDA) was used for exchange and correlation. The rotationally invariant LDA+U formalism was used and $U_{eff}$ was chosen as 6.52 eV for Cu. Electronic wavefunctions were expanded in plane waves with a kinetic energy cutoﬀ of 400 eV. The structures were relaxed until the energy and residual force on each atom were smaller than $10^{-4}$ eV and 0.02 eV ${\AA}^{-1}$, respectively. The k-points sampling was 16 ×16 ×1, generated automatically with the origin at the $\Upgamma$-point. The vacuum layer of the model was larger than 15 $\AA$.

\section{Result and discussion}

\begin{figure}[htbp] \centering
\includegraphics[width=0.8\textwidth]{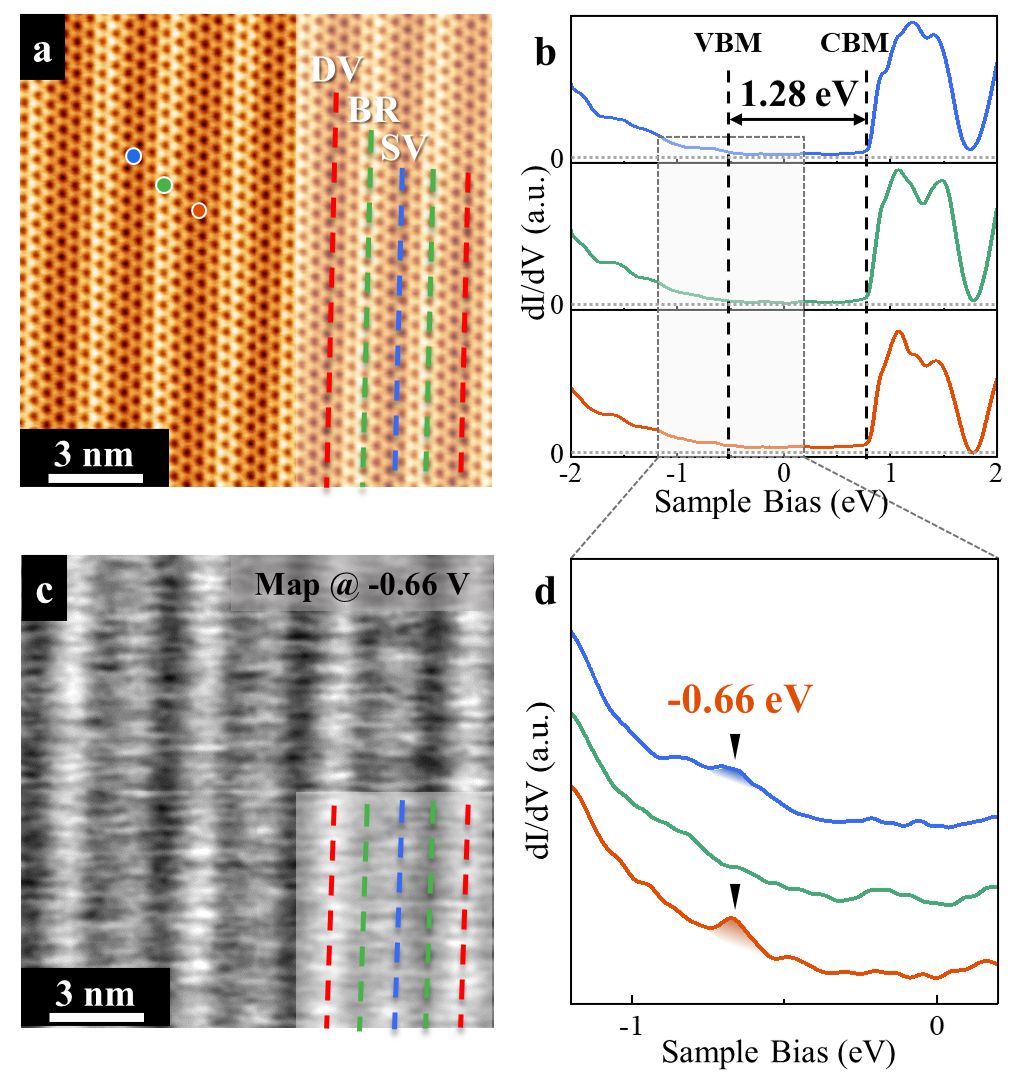}
    \caption{\textbf{Electronic properties of CuSe monolayer with 1D  moiré patterns.} (a) An atomically resolved STM image of 1D moiré patterns CuSe monolayer with hexagonal honeycomb lattice. (b) Three dI/dV curves collected at different positions which are marked by blue, green and red dots in (a). (c) A corresponding dI/dV map obtained at the energy of -0.66 V. Red, blue and green dashed lines in (a) and (c) indicate three different regions in 1D moiré patterns. (d) Zoom-in dI/dV curves from a gray square in (b). Two black triangles represent a peak at -0.66 V. Scanning parameters: (a) $V_s$ = -0.42 V, $I_t$ = 300 pA; (b) $V_s$ = 1.5 V, $I_t$ = 400 pA, $V_{rms}$ = 10 mV; (c) $V_s$ = -0.66 V, $I_t$ = 300 pA.}
    \label{fig;ele_prop}
\end{figure}

The CuSe monolayer with 1D moiré patterns is grown by a direct one-step selenization of a Cu(111) substrate at low Se coverage as described in our previously published papers\cite{RN11,RN12,RN14,RN16}. Due to the uniaxially strained honeycomb lattice of CuSe monolayer with respect to the Cu(111) substrate, large-scale and well-defined 1D moiré patterns are formed\cite{RN12}. Moreover, some mirror twin boundaries locate on the 1D moiré patterns of the CuSe monolayer, which will make the CuSe lattice at both ends of the line defects mirror-symmetrically arranged. (Fig. \ref{fig;S1}). Fig. \ref{fig;ele_prop}(a) shows an atomically resolved STM image of the CuSe monolayer endowed with a well arrangement 1D moiré patterns\cite{RN12,RN16}. Based on the space variation between bright stripes (see a line-profile in Figure S2(c)), CuSe surface could be divided into three regions: deep valley (DV), shallow valley (SV) and bright ridge (BR) regions, as shown in Fig. \ref{fig;ele_prop}(a). For clarity, the three regions are marked with three straight colored dashed lines in the lower right corner of Fig. \ref{fig;ele_prop}(a). To investigate the electronic properties of the CuSe monolayer, we then performed the STS characterization. figure S2(b) shows three waterfall plots of normalized dI/dV curves along the gradient arrows in the three distinct regions in Fig. \ref{fig;S2}(a), indicating a uniform electronic distribution in each region. Fig. \ref{fig;ele_prop}(b) shows three representative dI/dV curves collected at blue, green and red dots in three regions, respectively. All three curves show the same semiconducting behavior with valence band maximum (VBM) located at –0.52 eV and conductance band minimum (CBM) located at 0.76 eV, resulting an electronic bandgap of $\sim$1.28 eV. 

Fig. \ref{fig;ele_prop}(d) shows zoom-in dI/dV curves from Fig. \ref{fig;ele_prop}(b) with the energy range from -1.2 eV to 0.2 eV. Interestingly, the blue and red curves obtained at the SV and DV regions manifest a small peak at –0.66 eV, while the green curve from the BR region has no this feature. Additionally, the peak intensity for the DV region is slightly stronger than that for the SV regions. The peak intensity difference can be clearly identified in a dI/dV map at –0.66 eV (Fig. \ref{fig;ele_prop}(c)), showing an obvious 1D periodic modulation of the electronic structure for the CuSe monolayer on the Cu(111) substrate. To figure out the physical origin of the –0.66 eV peak, the projected density of states (PDOSs) in different regions are further calculated. As shown in Fig. \ref{fig;S3}, the –0.5 eV peaks are clearly observed in both blue and red curves in the calculated PDOSs contributed by in-plane orbitals (Se $p_x/p_y$ and Cu $d_{xy}/d_{x^2-y^2}/d_{z^2}$) (Figure S3(d)), while the green curve had no –0.5 eV peak. The trend of the –0.5 eV peaks in Fig. \ref{fig;S3}(d) is consistent with that of –0.66 eV peaks in dI/dV curves (Fig. \ref{fig;ele_prop}(c)), indicating that the –0.66 eV peak in dI/dV curves is mainly contributed by in-plane orbitals of CuSe monolayer at SV and DV regions.

\begin{figure}
\centering
\includegraphics[width=0.8\textwidth]{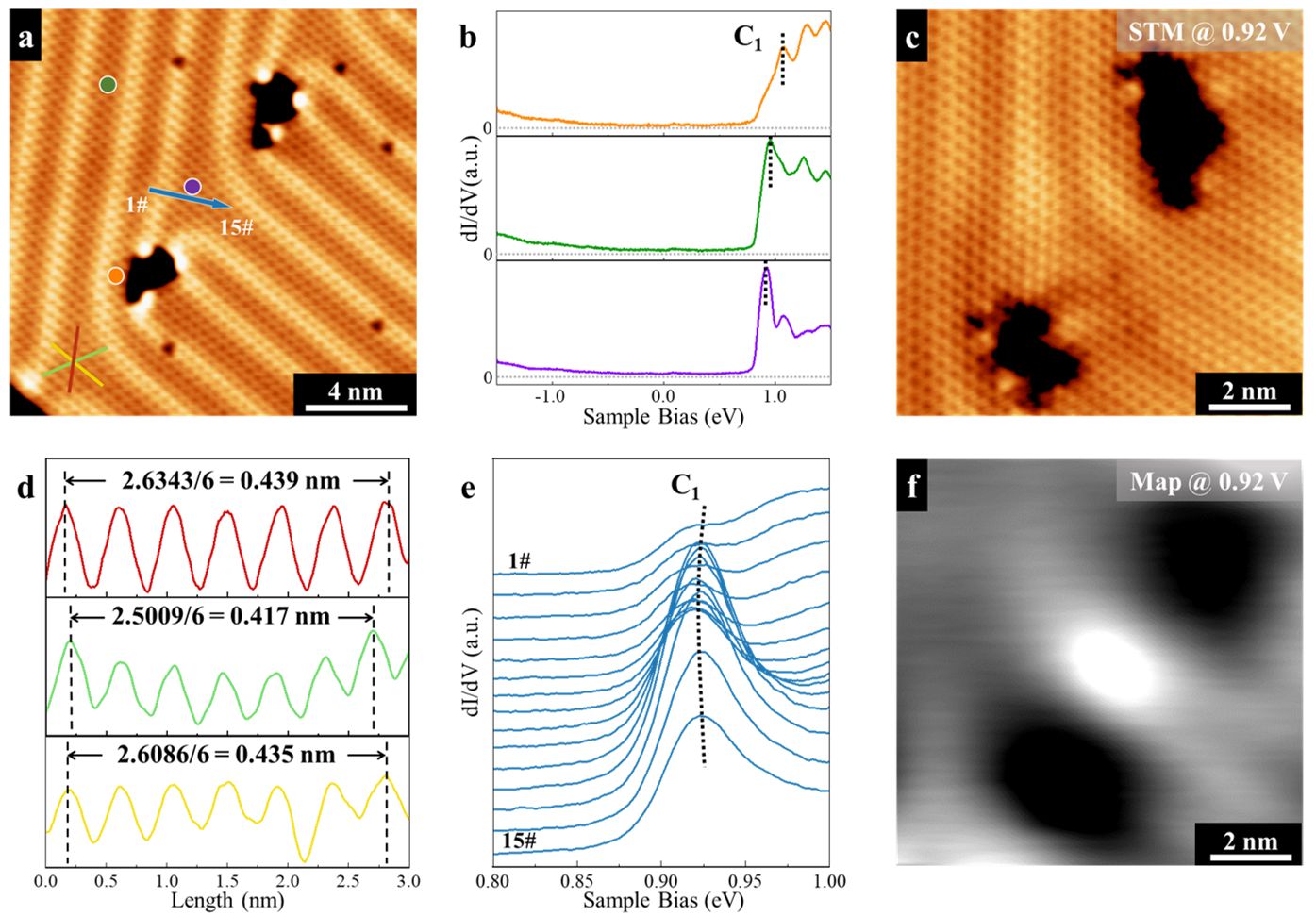}
    \caption{\textbf{Atomic structures and electronic properties of domain boundary of CuSe monolayer on Cu(111) substrate.} (a) A high resolution STM image of the domain boundary. (b) dI/dV curves collected at three positions, as indicated by colored dots in (a). Black dashed lines indicate the positions of C1 peak in the conduction band. (c) and (f) STM image and corresponding dI/dV map at the energy of 0.92 V. (d) Three line-profiles at the domain boundary across the red, green, and yellow lines in (a). (e) Waterfall plots of  dI/dV curves along a blue arrow in (a). A black curved dashed line indicates the C1 peak movement. Scanning parameters: (a) $V_s$ = 50 mV, $I_t$ = 1.3 nA; (b) $V_s$ = 2 V, $I_t$ = 300 pA, $V_{rms}$ = 10 mV; (c) $V_s$ = 0.92 V, $I_t$ = 300 pA; (e) $V_s$ = 2 V, $I_t$ = 300 pA, $V_{rms}$ = 10 mV; (f) $V_s$ = 0.92 mV, $I_t$ = 300 pA, $V_{rms}$ = 10 mV.} 
    \label{fig;strct} 
\end{figure}

Due to the 6-fold symmetry of Cu(111) substrate, the as-synthesized 1D moiré patterns in the CuSe monolayer manifest three equivalent domains\cite{RN12}. Fig. \ref{fig;strct}(a) depicts a typical STM image showing a boundary formed by two different orientations of 1D moiré patterns. Interestingly, the partial connection area between two domains shows regularly arranged and continuous six-membered rings (see the high resolution STM image for the domain boundary in Fig. \ref{fig;S4}(b)). Fig. \ref{fig;strct}(d) shows three line-profiles along three directions of CuSe monolayer at the domain boundary as indicated by red, green and yellow dashed lines in Fig. \ref{fig;strct}(a), where the directions of the red and yellow dashed lines are parallel to the orientation of 1D moiré pattern of two domains on both sides of the boundary, respectively. The measured periodicity of CuSe honeycomb lattice in three directions is 0.439 nm (red curve), 0.417 nm (green curve), and 0.435 nm (yellow curve). Obviously, the periodicity in the red and yellow directions is very similar and significantly larger than that in the green direction. We then confirm that the lattice structure of CuSe monolayer in this area of the domain boundary is stretched by a biaxial strain. However, the domain boundary also exists few large irregularly shaped defects shown in Fig. \ref{fig;strct}(a). This confirms that the perfect connection between two domains of CuSe monolayer will produce great strain accumulation at the boundary, thus resulting in the formation of large defects. A convincing evidence is that the lattice constant around the irregularly shaped defect in Fig. \ref{fig;S4}(c) (0.406 nm) is slightly smaller than that in the internal biaxial strain (BS) (0.435 nm in Fig. \ref{fig;strct}(d)). It should be noted that thanks to the formed large defects in the domain boundary, the CuSe honeycomb lattice near the large defect will be not affected by strain.

To investigate the effect of strain on the electronic structures of CuSe monolayer, we carried out the differential conductance spectroscopy (dI/dV) measurements. Fig. \ref{fig;strct}(b) shows three dI/dV curves collected from three distinct regions in the Fig. \ref{fig;strct}(a). The orange curve taken near the edge of a large irregularly shaped defect represents the electronic properties of CuSe monolayer without strain, while the green curve collected from the DV region reflects the electronic properties of CuSe monolayer distorted by uniaxial strain (US), and the purple curve obtained from the domain boundary gives access to the electronic properties of CuSe monolayer distorted by biaxial strain (BS). Although the three curves show similar semiconducting behavior, the first peak labelled by $C_1$ in conduction band has an obvious movement, as marked by a series of black dashed lines. The $C_1$ peak locates at 1.06 eV for the strain-free region, while for the US region and BS region, this peak shifts down to 0.95 eV and 0.92 eV, respectively (see Fig. \ref{fig;S5} for more dI/dV curves near the strain-free region). Additionally, in order to further investigate the movement of $C_1$ peak, we took a series of dI/dV spectra along a blue arrow from one 1D moiré pattern domain to another domain via domain boundary in Fig. \ref{fig;strct}(a), where the stain in the CuSe monolayer changes from US vis BS to US. Fig. \ref{fig;strct}(e) shows waterfall plots of normalized dI/dV curves, and a black curved dashed line describes the changing trend of $C_1$ peak in conduction band. Evidently, the $C_1$ peak slightly moves down from 0.93 eV to 0.92 eV and then gradually increases to 0.93 eV, which further confirms the strain can modulate the electronic structures of CuSe monolayer on the Cu(111) substrate. Fig. \ref{fig;strct}(c) and \ref{fig;strct}(f) are STM image around the domain boundary and its corresponding dI/dV map at 0.92 eV, respectively. It can be clearly found that the electronic states are strongly concentrated in the BS area (domain boundary) at 0.92 eV, in good agreement with STS results in Fig. \ref{fig;strct}(b) and \ref{fig;strct}(e).

\begin{figure}
\centering
\includegraphics[width=0.6\textwidth]{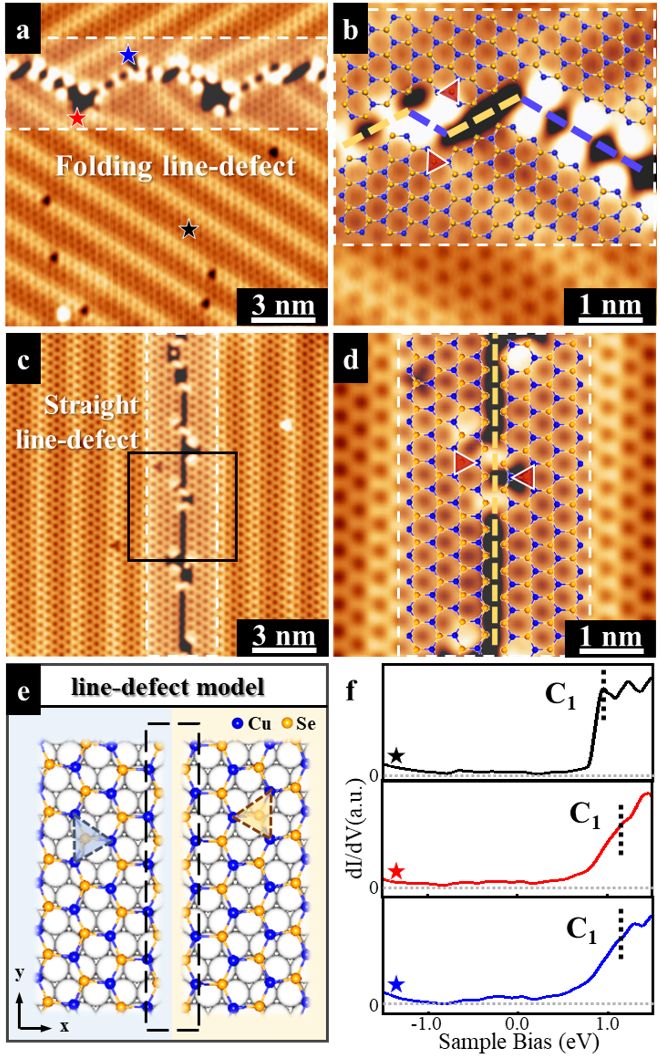}
    \caption{\textbf{Line defects in the CuSe monolayer on the Cu(111) substrate.} (a, b) A large-scale STM image and corresponding high-resolution STM image of the folding line-defect. (c, d) A large-scale STM image and corresponding high-resolution STM image of the straight line-defect. Atomic models are covered into (b) and (d) to shed light on the edge terminations. (e) Schematic description of straight line-defect in CuSe atomic model, where blue and yellow triangles indicate that the left and right models are two mirror-symmetric domains. (f) dI/dV curves taken at three positions, as marked by black, red and blue stars in (a). Scanning parameters: (a) $V_s$ = 0.4 V, $I_t$ = 200 pA; (b) $V_s$ = 2 V, $I_t$ = 300 pA; (c) $V_s$ = 0.2 V, $I_t$ = 100 pA; (d) $V_s$ = 0.2 V, $I_t$ = 100 pA; (f) $V_s$ = 2 V, $I_t$ = 400 pA, $V_{rms}$ = 10 mV.}
    \label{fig;lin-def}
\end{figure}

In addition to the domain boundaries consisted of two different orientations of 1D moiré patterns of CuSe monolayer on the Cu(111) substrate, we also observed two kinds of line defects formed by mirror-symmetric domains on both sides (Fig. \ref{fig;lin-def}(a) and \ref{fig;lin-def}(c)). Fig. \ref{fig;lin-def}(b) and \ref{fig;lin-def}(d) are zoom-in STM images of Fig. \ref{fig;lin-def}(a) and \ref{fig;lin-def}(c), showing atomically resolved structures of the line defects, respectively. A schematic description of edge terminations of a CuSe hexagonal island is illustrated in Fig. \ref{fig;S6}, where the Cu-edge and Se-edge are represented by blue and yellow dashed lines, respectively. Fig. \ref{fig;lin-def}(e) presents a schematic diagram of the line defect formed by two mirror-symmetric CuSe monolayer domains, where the mirror domains can be clearly identified by blue and yellow triangles. To identify the edge terminations of the observed line defects, we superimpose the atomic models of two mirror domains of CuSe monolayer into Fig. \ref{fig;lin-def}(b) and \ref{fig;lin-def}(d). The results show that the edge terminations are an alternate Cu-edge and Se edge in Fig. \ref{fig;lin-def}(b), forming a folding line defect, while Fig. \ref{fig;lin-def}(d) presents a straight Se edge, forming a straight-line defect. In order to study the electronic properties of CuSe monolayer near the line defect, where the lattice structure should be free of strain, we carried out dI/dV measurements. Fig. \ref{fig;lin-def}(f) shows STS curves taken near the line defect and terrace of CuSe monolayer, as marked by black, red and blue stars in Fig. \ref{fig;lin-def}(a). The STS results clearly show the movement of the $C_1$ peak positions, as indicated by black dashed lines in Fig. \ref{fig;lin-def}(g), further confirming that the strain will induce the peak position to move towards downward. In addition, due to the existence of edge dangling bonds outside the defect, the freestanding CuSe has trivial and non-trivial boundary states\cite{RN12}. However, on the Cu(111) substrate, the edge dangling bond of CuSe is saturated by the Cu(111) substrate, so there is no doubt that it doesn’t show the boundary state externally (Fig. \ref{fig;S7}).

\begin{figure}
\centering
\includegraphics[width=\textwidth]{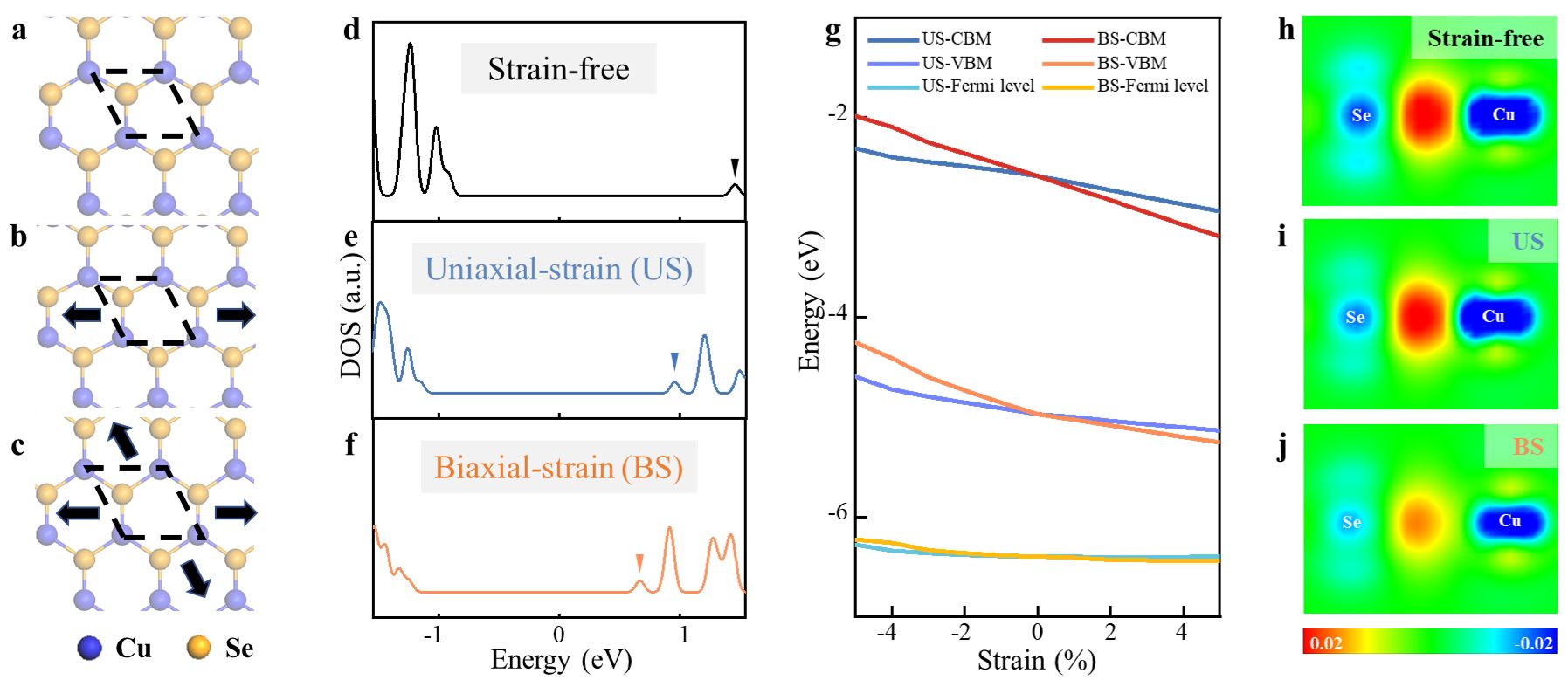}
    \caption{\textbf{Atomic configurations and calculated electronic structures of CuSe monolayer.} (a-c) Atomic configurations of CuSe monolayer under strain-free, 7.3\% uniaxial-strain and 7.3\% biaxial-strain, respectively. The primitive cell of CuSe monolayer is denoted by the black dotted line. (d-f) DOSs of CuSe monolayer under strain-free, 7.3\% uniaxial-strain and 7.3\% biaxial-strain, respectively. (g) CBM, VBM and Fermi level variations under uniaxial-strain and biaxial-strain of CuSe monolayer. (h-j) Cross sections of differential charge density (e/Bohr3) of CuSe monolayer under strain-free, 7.3\% uniaxial-strain and 7.3\% biaxial-strain, respectively.}
    \label{fig;DOS}
\end{figure}

In order to further investigate the electronic properties of monolayer CuSe, we performed first-principles calculations on (1 × 1) CuSe monolayer under strain-free, uniaxial strain (US) and biaxial strain (BS) respectively, as shown in Fig. \ref{fig;DOS}(a)-\ref{fig;DOS}(c). The magnitude of strain is defined as $\varepsilon$= ($a$-$a_0$)/$a_0$, where $a$ and $a_0$ are the lattice constants of the strained and unstrained structures, respectively. According to the published paper, the lattice constants of CuSe monolayer along the orientation of 1D moiré pattern is 7.3\% larger than those along other directions\cite{RN12}. The density of states (DOSs) of CuSe monolayer under strain-free condition is shown in Fig. \ref{fig;DOS}(d). The first peak ($C_1$) above the Fermi level is about 1.44 eV. Under 7.3\% US and 7.3\% BS, the $C_1$ peak shifts left to 0.95 eV (Fig. \ref{fig;DOS}(e)) and 0.66 eV (Fig. \ref{fig;DOS}(f)) respectively, which is in good agreement with the experimental dI/dV curves in Fig. \ref{fig;strct}(b).

To better understand how strain affects the electronic structures of CuSe monolayer, first-principles calculations of CuSe monolayer with US and BS ranging from -5\% to 5\% were further performed. As shown in Fig. \ref{fig;DOS}(g), the CBM, VBM and Fermi level variations with BS of CuSe monolayer are larger than those with US. Under both US and BS, the CBM and VBM of CuSe monolayer move more with the increasing strain. Furthermore, the differential charge densities of CuSe monolayer under strain-free, 7.3\% US and 7.3\% BS were also investigated. The differential charge densities of CuSe monolayer under strain-free (Fig. \ref{fig;DOS}(h)) shows charge accumulations between Cu and Se atoms, while charge depletion on Cu and Se atoms, hence demonstrating a covalent bonding character. Under 7.3\% US (Fig. \ref{fig;DOS}(i) and 7.3\% BS (Fig. \ref{fig;DOS}(j)), the charge accumulations between Cu and Se atoms and the charge depletions on Cu and Se atoms become less, demonstrating weaker covalent bonding characters which is the main cause of the $C_1$ peak shift to the left. To figure out the physical origin of the $C_1$ peak, the PDOSs of monolayer CuSe under strain-free, 7.3\% uniaxial-strain and 7.3\% biaxial-strain on Cu(111) are further calculated. As shown in Fig. \ref{fig;S8}, the $C_1$ peak is mainly dominated by s orbitals of CuSe monolayer.

\section{Conclusion}

In summary, we demonstrate the direct investigation of electronic properties of CuSe monolayer on the Cu(111) substrate under different strain conditions, fully characterized by combined STM/STS measurements, and supported by DFT calculations. The uniaxial strain induced 1D moiré patterns produces 1D periodic modulation of electronic structures of CuSe monolayer. Additionally, compared to the uniaxial strain region at the terrace, electronic properties of CuSe monolayer at biaxial strain region existed in the domain boundary and strain-free region found near the line defects show that the first peak in conduction band has a tendency to move in the opposite direction, which has been successfully confirmed by DFT calculations. The results not only enrich the fundamental understanding toward the role of strain in modulating electronic properties of CuSe monolayer, but also lay the foundation for the development of 2D semiconductor materials.

\section{Acknowledgement}

This work was supported by the National Natural Science Foundation of China (Nos. 62271238 and 61901200), the Yunnan Fundamental Research Projects (Nos. 202201AT070078, 202101AV070008, 202101AW070010, and 202101AU070043), the Strategic Priority Research Program of Chinese Academy of Sciences (XDB30000000), the Analysis and Testing Foundation of KUST (2021T20170056), and the Dongguan Innovation Research Team Program. Numerical computations were performed on Hefei advanced computing center.

\bibliographystyle{unsrt}
\bibliography{ref}

\begin{thebibliography}{10}

\bibitem{RN1}
K.~F. Mak, K.~L. McGill, J.~Park, and P.~L. McEuen.
\newblock The valley hall effect in mos2 transistors.
\newblock {\em Science}, 344(6191):1489--1492, 2014.

\bibitem{RN2}
N.~Mounet, M.~Gibertini, P.~Schwaller, D.~Campi, A.~Merkys, A.~Marrazzo,
  T.~Sohier, I.~E. Castelli, A.~Cepellotti, G.~Pizzi, and N.~Marzari.
\newblock Two-dimensional materials from high-throughput computational
  exfoliation of experimentally known compounds.
\newblock {\em Nat. Nanotech.}, 13(3):246--+, 2018.

\bibitem{RN3}
S.~Y. Zhu, Y.~Shao, E.~Wang, L.~Cao, X.~Y. Li, Z.~L. Liu, C.~Liu, L.~W. Liu,
  J.~O. Wang, K.~Ibrahim, J.~T. Sun, Y.~L. Wang, S.~X. Du, and H.~J. Gao.
\newblock Evidence of topological edge states in buckled antimonene monolayers.
\newblock {\em Nano Lett.}, 19(9):6323--6329, 2019.

\bibitem{RN4}
Zhican Zhou, Fengyou Yang, Shu Wang, Lei Wang, Xiaofeng Wang, Cong Wang, Yong
  Xie, and Qian Liu.
\newblock Emerging of two-dimensional materials in novel memristor.
\newblock {\em Front. Phys.}, 17(2):23204, 2021.

\bibitem{RN5}
Ya-Hui Mao, Huan Shan, Jin-Rong Wu, Ze-Jun Li, Chang-Zheng Wu, Xiao-Fang Zhai,
  Ai-Di Zhao, and Bing Wang.
\newblock Observation of pseudogap in snse2 atomic layers grown on graphite.
\newblock {\em Front. Phys.}, 15(4):43501, 2020.

\bibitem{RN6}
Jiang-Xiazi Lin, Ya-Hui Zhang, Erin Morissette, Zhi Wang, Song Liu, Daniel
  Rhodes, K~Watanabe, T~Taniguchi, James Hone, and Jia Li.
\newblock Spin-orbit–driven ferromagnetism at half moiré filling in
  magic-angle twisted bilayer graphene.
\newblock {\em Science}, 375(6579):437--441, 2022.

\bibitem{RN7}
X.~Y. Wang, H.~Zhang, Z.~L. Ruan, Z.~L. Hao, X.~T. Yang, J.~M. Cai, and J.~C.
  Lu.
\newblock Research progress of monolayer two-dimensional atomic crystal
  materials grown by molecular beam epitaxy in ultra-high vacuum conditions.
\newblock {\em Acta Phys. Sin-Ch Ed}, 69(11):118101, 2020.

\bibitem{RN8}
Bing Liu, Jian Liu, Guangyao Miao, Siwei Xue, Shuyuan Zhang, Lixia Liu,
  Xiaochun Huang, Xuetao Zhu, Sheng Meng, Jiandong Guo, Miao Liu, and Weihua
  Wang.
\newblock Flat agte honeycomb monolayer on ag(111).
\newblock {\em J. Phys. Chem. Lett.}, 10(8):1866--1871, 2019.

\bibitem{RN9}
J.~Shah, H.~M. Sohail, R.~I.~G. Uhrberg, and W.~Wang.
\newblock Two-dimensional binary honeycomb layer formed by ag and te on
  ag(111).
\newblock {\em J. Phys. Chem. Lett.}, 11(5):1609--1613, 2020.

\bibitem{RN10}
Maximilian Ünzelmann, Hendrik Bentmann, Philipp Eck, Tilman Kißlinger,
  Begmuhammet Geldiyev, Janek Rieger, Simon Moser, Raphael~C. Vidal, Katharina
  Kißner, Lutz Hammer, M.~Alexander Schneider, Thomas Fauster, Giorgio
  Sangiovanni, Domenico Di~Sante, and Friedrich Reinert.
\newblock Orbital-driven rashba effect in a binary honeycomb monolayer agte.
\newblock {\em Phys. Rev. Lett.}, 124(17):176401, 2020.

\bibitem{RN11}
X.~Lin, J.~C. Lu, Y.~Shao, Y.~Y. Zhang, X.~Wu, J.~B. Pan, L.~Gao, S.~Y. Zhu,
  K.~Qian, Y.~F. Zhang, D.~L. Bao, L.~F. Li, Y.~Q. Wang, Z.~L. Liu, J.~T. Sun,
  T.~Lei, C.~Liu, J.~O. Wang, K.~Ibrahim, D.~N. Leonard, W.~Zhou, H.~M. Guo,
  Y.~L. Wang, S.~X. Du, S.~T. Pantelides, and H.~J. Gao.
\newblock Intrinsically patterned two-dimensional materials for selective
  adsorption of molecules and nanoclusters.
\newblock {\em Nat. Mater.}, 16(7):717--721, 2017.

\bibitem{RN12}
Lei Gao, Jia-Tao Sun, Jian-Chen Lu, Hang Li, Kai Qian, Shuai Zhang, Yu-Yang
  Zhang, Tian Qian, Hong Ding, Xiao Lin, Shixuan Du, and Hong-Jun Gao.
\newblock Epitaxial growth of honeycomb monolayer cuse with dirac nodal line
  fermions.
\newblock {\em Adv. Mater.}, 30(16):1707055, 2018.

\bibitem{RN13}
Jianchen Lu, Lei Gao, Shiru Song, Hang Li, Gefei Niu, Hui Chen, Tian Qian, Hong
  Ding, Xiao Lin, Shixuan Du, and Hong-Jun Gao.
\newblock Honeycomb agse monolayer nanosheets for studying two-dimensional
  dirac nodal line fermions.
\newblock {\em ACS Appl. Nano Mater.}, 4(9):8845--8850, 2021.

\bibitem{RN14}
Xingyue Wang, Zilin Ruan, Renjun Du, Hui Zhang, Xiaotian Yang, Gefei Niu,
  Jinming Cai, and Jianchen Lu.
\newblock Structural characterizations and electronic properties of cuse
  monolayer endowed with triangular nanopores.
\newblock {\em J. Mater. Sci.}, 56(17):10406--10413, 2021.

\bibitem{RN15}
Lei Gao, Yan-Fang Zhang, Jia-Tao Sun, and Shixuan Du.
\newblock Band engineering of honeycomb monolayer cuse via atomic modification.
\newblock {\em Chin. Phys. B}, 30(10):106807, 2021.

\bibitem{RN16}
Gefei Niu, Jianchen Lu, Xingyue Wang, Zilin Ruan, Hui Zhang, Lei Gao, Jinming
  Cai, and Xiao Lin.
\newblock Se-concentration dependent superstructure transformations of cuse
  monolayer on cu(111) substrate.
\newblock {\em 2D Mater.}, 9(1):015017, 2021.

\bibitem{RN17}
Zhipeng Song, Jierui Huang, Shuai Zhang, Yun Cao, Chen Liu, Ruizi Zhang,
  Qi~Zheng, Lu~Cao, Li~Huang, Jiaou Wang, Tian Qian, Hong Ding, Wu~Zhou,
  Yu-Yang Zhang, Hongliang Lu, Chengmin Shen, Xiao Lin, Shixuan Du, and
  Hong-Jun Gao.
\newblock Observation of an incommensurate charge density wave in monolayer
  tise2/cuse/cu(111) heterostructure.
\newblock {\em Phys. Rev. Lett.}, 128(2):026401, 2022.

\bibitem{RN18}
Kewei Tang and Weihong Qi.
\newblock Moiré-pattern-tuned electronic structures of van der waals
  heterostructures.
\newblock {\em Adv. Func. Mater.}, 30(32):2002672, 2020.

\bibitem{RN19}
Jun Kang, Jingbo Li, Shu-Shen Li, Jian-Bai Xia, and Lin-Wang Wang.
\newblock Electronic structural moire pattern effects on mos2/mose2 2d
  heterostructures.
\newblock {\em Nano Lett.}, 13(11):5485--5490, 2013.

\bibitem{RN20}
Kyle~L. Seyler, Pasqual Rivera, Hongyi Yu, Nathan~P. Wilson, Essance~L. Ray,
  David~G. Mandrus, Jiaqiang Yan, Wang Yao, and Xiaodong Xu.
\newblock Signatures of moire-trapped valley excitons in mose2/wse2
  heterobilayers.
\newblock {\em Nature}, 567(7746):66--70, 2019.

\bibitem{RN21}
Chendong Zhang, Ming-Yang Li, Jerry Tersoff, Yimo Han, Yushan Su, Lain-Jong Li,
  David~A. Muller, and Chih-Kang Shih.
\newblock Strain distributions and their influence on electronic structures of
  wse2-mos2 laterally strained heterojunctions.
\newblock {\em Nat. Nanotechnol.}, 13(2):152--158, 2018.

\bibitem{RN22}
Yi~Pan, Haigang Zhang, Dongxia Shi, Jiatao Sun, Shixuan Du, Feng Liu, and
  Hong-jun Gao.
\newblock Highly ordered, millimeter-scale, continuous, single-crystalline
  graphene monolayer formed on ru (0001).
\newblock {\em Adv. Mater.}, 21(27):2777--2780, 2009.

\bibitem{RN23}
Y.~S. Bai, L.~Zhou, J.~Wang, W.~J. Wu, L.~J. McGilly, D.~Halbertal, C.~F.~B.
  Lo, F.~Liu, J.~Ardelean, P.~Rivera, N.~R. Finney, X.~C. Yang, D.~N. Basov,
  W.~Yao, X.~D. Xu, J.~Hone, A.~N. Pasupathy, and X.~Y. Zhu.
\newblock Excitons in strain-induced one-dimensional moire potentials at
  transition metal dichalcogenide heterojunctions.
\newblock {\em Nat. Mater.}, 19(10):1068--1073, 2020.

\bibitem{RN24}
Q.~J. Tong, H.~Y. Yu, Q.~Z. Zhu, Y.~Wang, X.~D. Xu, and A.~Yao.
\newblock Topological mosaics in moire superlattices of van der waals
  heterobilayers.
\newblock {\em Nat. Phys.}, 13(4):356--362, 2017.

\bibitem{RN25}
I.~Horcas, R.~Fernandez, J.~M. Gomez-Rodriguez, J.~Colchero, J.~Gomez-Herrero,
  and A.~M. Baro.
\newblock Wsxm: A software for scanning probe microscopy and a tool for
  nanotechnology.
\newblock {\em Rev. Sci. Instrum.}, 78(1):8, 2007.

\end{thebibliography}

\pagebreak
\begin{center}
\title{\huge\textbf{Supporting Information} \protect\\
\setlength{\parskip}{2em}
\LARGE\textbf{Electronic properties of monolayer copper selenide with one-dimensional moiré patterns}}\\
\setlength{\parskip}{2em}
\large{\author{Gefei Niu$^{1\#}$, Jianchen Lu$^{1\#}$, Jianqun Geng$^{1\#}$, Shicheng Li$^1$, Hui Zhang$^1$, Wei Xiong$^1$, \\Zilin Ruan$^1$, Yong Zhang$^1$, Boyu Fu$^1$, Lei Gao$^{2\ast}$, Jinming Cai$^{1\ast}$}\\
\setlength{\parskip}{2em}
\date{%
    $^1$Faculty of Materials Science and Engineering, Kunming University of Science and Technology, No. 68 Wenchang Road, Kunming 650093, China.\\
    $^2$Faculty of Science, Kunming University of Science and Technology, No. 727 Jingming South Road, Kunming 650500, China.
    \leftline{$^{\#}$ These authors contributed equally.}
    \leftline{$^{\ast}$ Corresponding Authors: lgao@kust.edu.cn, jclu@kust.edu.cn, j.cai@kust.edu.cn}
}
}

\end{center}

\setcounter{equation}{0}
\setcounter{figure}{0}
\setcounter{table}{0}
\setcounter{page}{1}
\makeatletter
\renewcommand{\theequation}{S\arabic{equation}}
\renewcommand{\thefigure}{S\arabic{figure}}

\begin{figure}[H]
\centering
\includegraphics[width=0.6\textwidth]{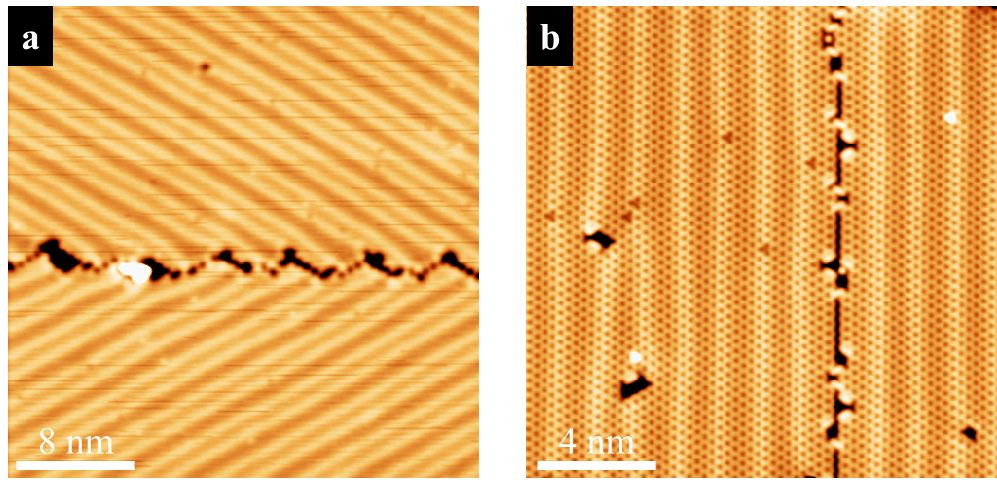}
    \caption{\textbf{Morphology of line-defect formed by mirror-symmetric domains.} (a) A large-scale STM image of the folding line-defect. (b) A large-scale STM image of the straight line-defect. Scanning parameters: (a) $V_s$ = 2 V, $I_t$ = 50 pA; (b) $V_s$ = 0.2 V, $I_t$ = 100 pA.}
    \label{fig;S1}
\end{figure}

\begin{figure}[H]
\centering
\includegraphics[width=0.6\textwidth]{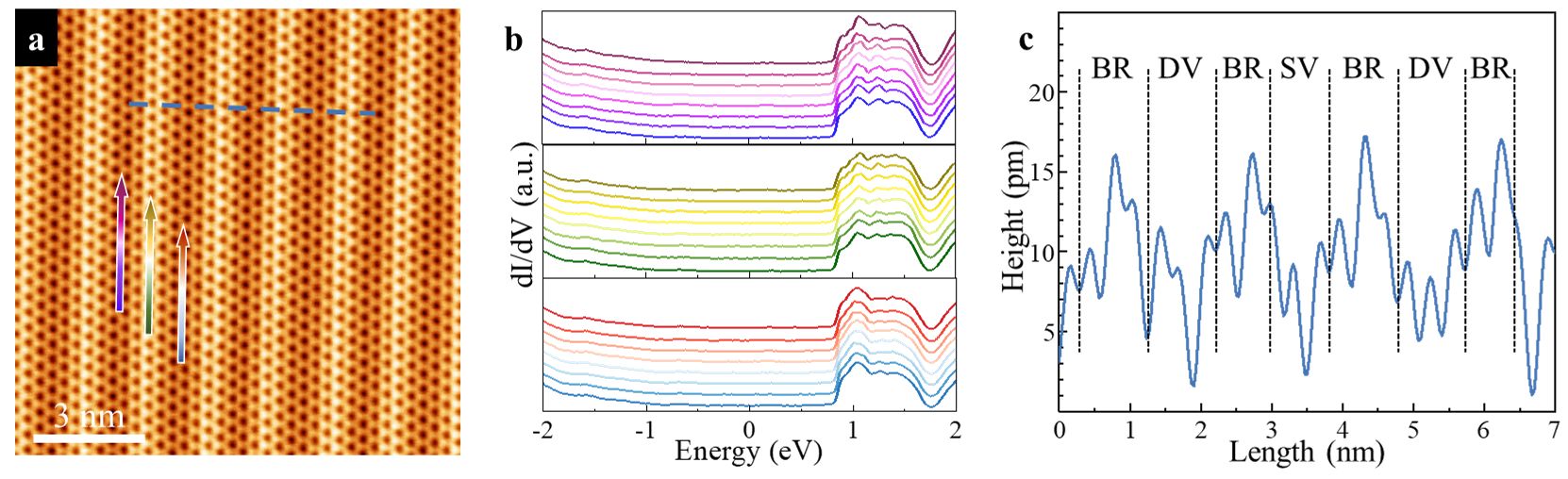}
    \caption{\textbf{Electronic properties of 1D moiré patterns of CuSe monolayer.} (a) An atomic-resolved STM image of 1D moiré patterns of CuSe monolayer with hexagonal honeycomb lattice. (b) Three waterfall plots of dI/dV curves along the gradient arrows in (a). (c) Line-profile perpendicular to 1D moiré patterns, indicated by a blue dashed line in (a), showing three distinct regions. Scanning parameters: (a) $V_s$ = -0.42 V, $I_t$ = 300 pA; (b) $V_s$ = 1.5 V, $I_t$ = 400 pA, $V_{rms}$ = 10 mV.}
    \label{fig;S2}
\end{figure}

\begin{figure}[H]
\centering
\includegraphics[width=0.6\textwidth]{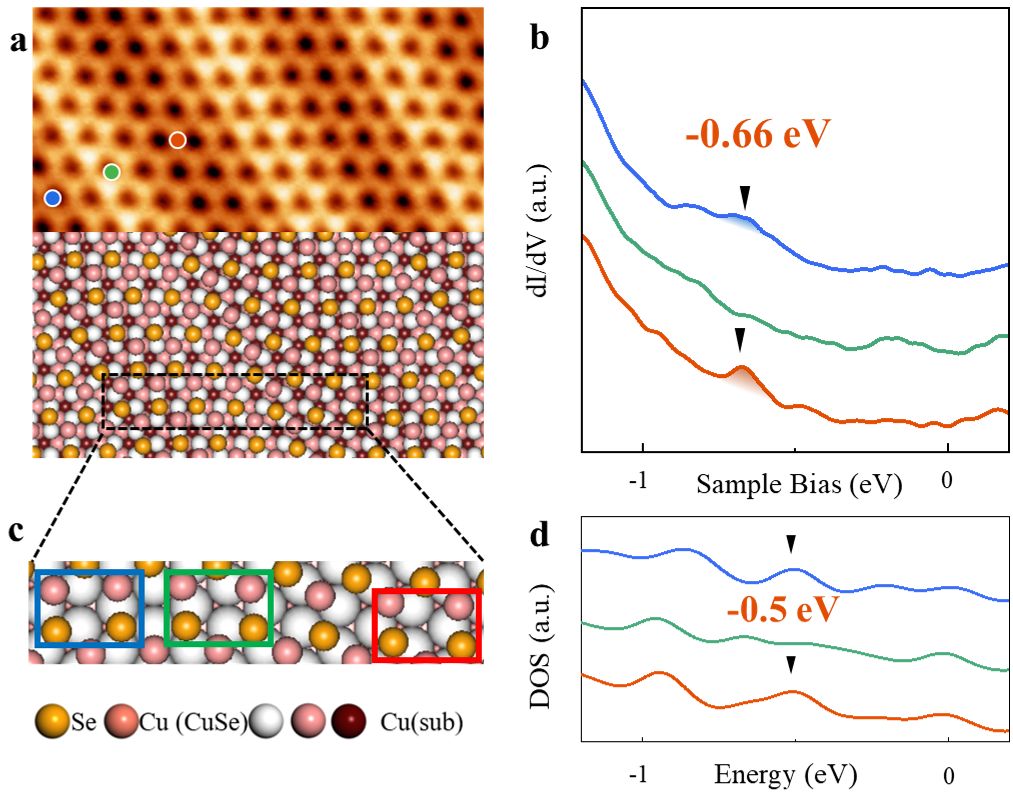}
    \caption{\textbf{Structural and electronic properties of monolayer CuSe on Cu(111).} (a) An atomically resolved STM image and corresponding optimized atomic structure of monolayer CuSe on Cu(111). (b) Three dI/dV curves collected at different positions which are marked by blue, green and red dots in (a). (c) Zoom-in atomic structure from a black square in (a). (d) Projected density of states (PDOSs) contributed by in-plane orbitals (Se $p_x/p_y$ and Cu $d_{xy}/d_{x^2-y^2}/d_{z^2}$) calculated at different positions which are marked by blue, green and red squares in (c).}
    \label{fig;S3}
\end{figure}

\begin{figure}[H]
\centering
\includegraphics[width=0.6\textwidth]{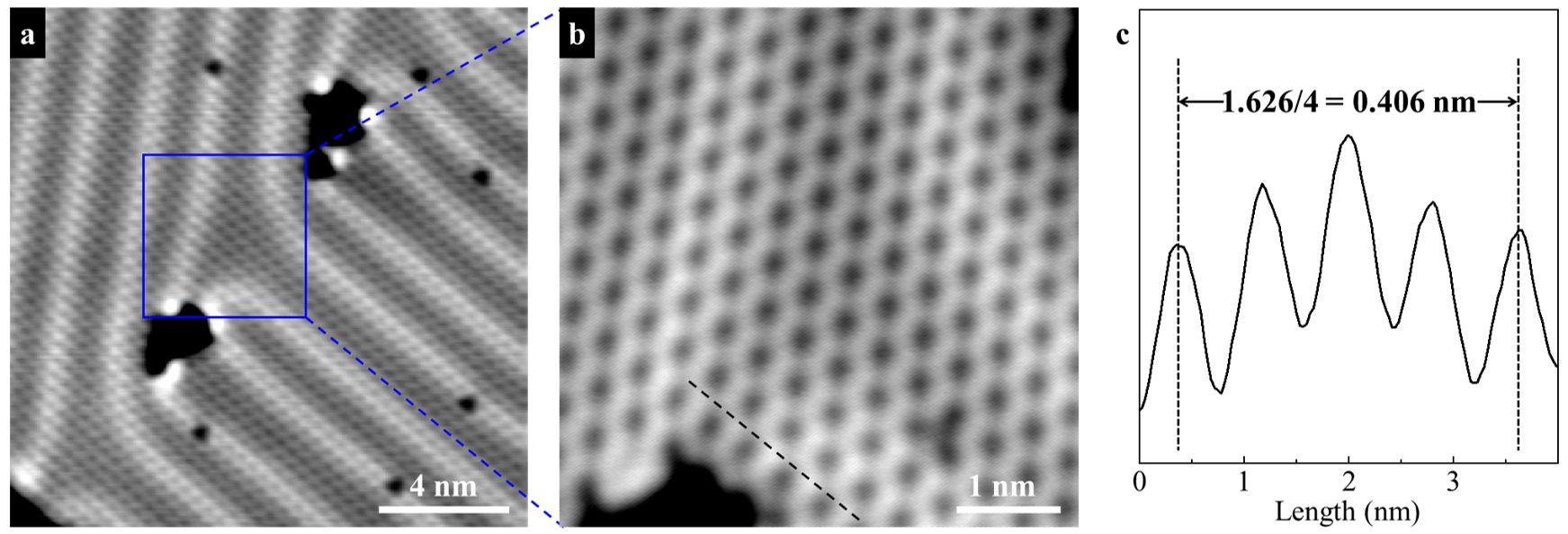}
    \caption{\textbf{Domain boundaries of CuSe monolayer with 1D moiré patterns.} (a) A large scale STM image of domain boundary of CuSe monolayer. (b) A zoom-in atomically resolved STM image of domain boundary, showing a continuous CuSe honeycomb lattice. (c) The line-profiles perpendicular to continuous boundary of 1D moiré patterns structure of CuSe, indicated by the black line in (b). Scanning parameters: (a) $V_s$ = 50 mV, $I_t$ = 100 pA; (b) $V_s$ = 10 mV, $I_t$ = 500 pA.}
    \label{fig;S4}
\end{figure}

\begin{figure}[H]
\centering
\includegraphics[width=0.6\textwidth]{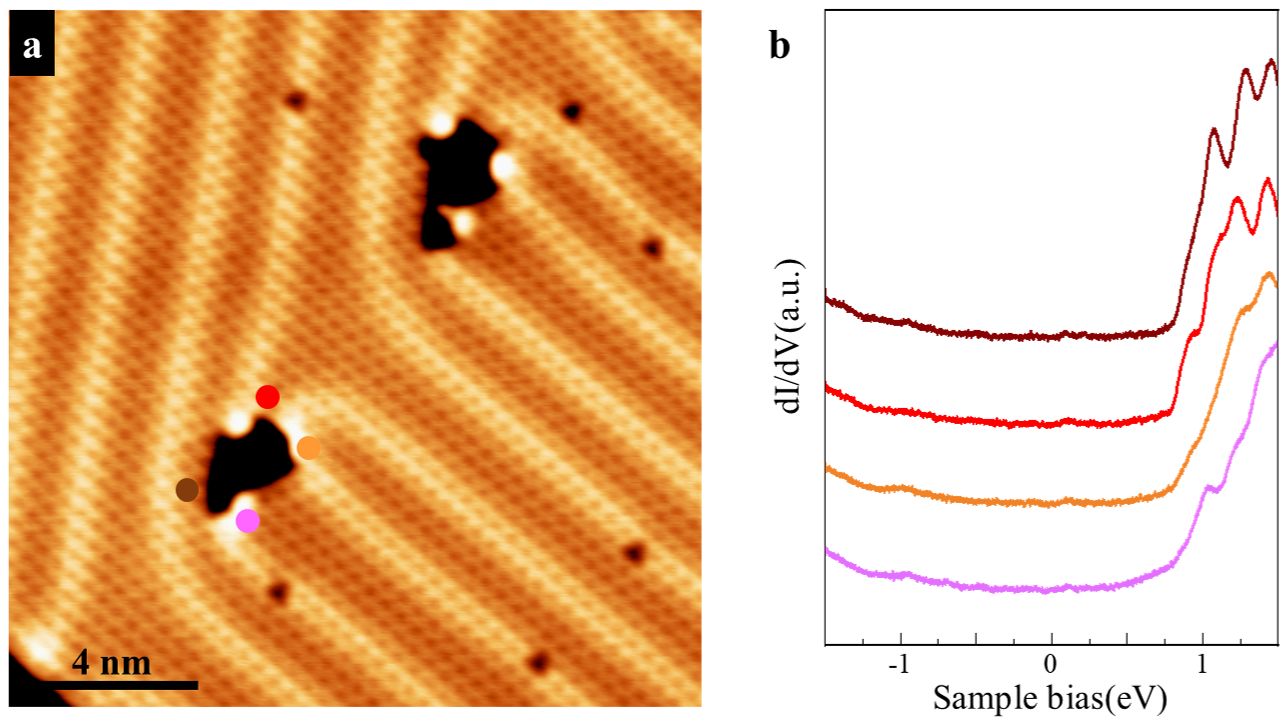}
    \caption{\textbf{The electronic properties around the defect of continuous boundaries.} (a) A large scale STM image of continuous boundary of CuSe monolayer with hexagonal honeycomb lattice. (b) dI/dV curves collected at four positions, as indicated by colored dots in (a). Scanning parameters: (a) Vs = 50 mV, It = 1.3 nA; (b) Vs = 2 V, It = 300 pA, Vrms = 10 mV.}
    \label{fig;S5}
\end{figure}

\begin{figure}[H]
\centering
\includegraphics[width=0.6\textwidth]{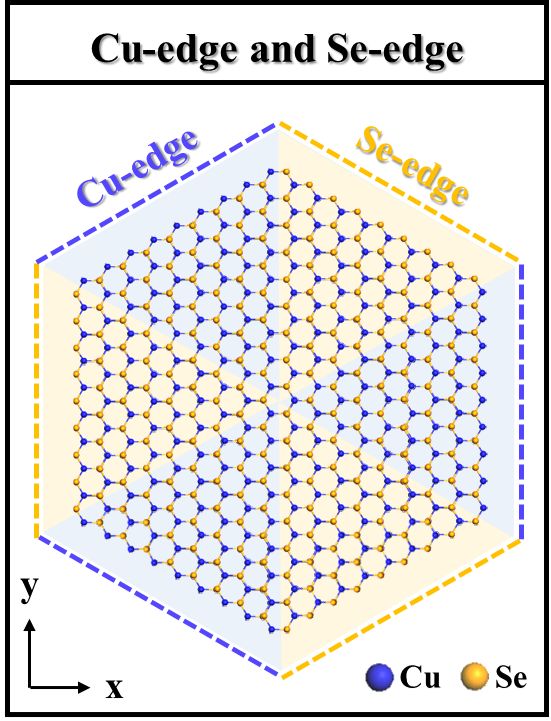}
    \caption{Schematic model of edge terminations of an hexagonal CuSe island, where the Cu edge and Se edge are illustrated by blue and yellow dashed lines, respectively.}
    \label{fig;S6}
\end{figure}

\begin{figure}[H]
\centering
\includegraphics[width=0.6\textwidth]{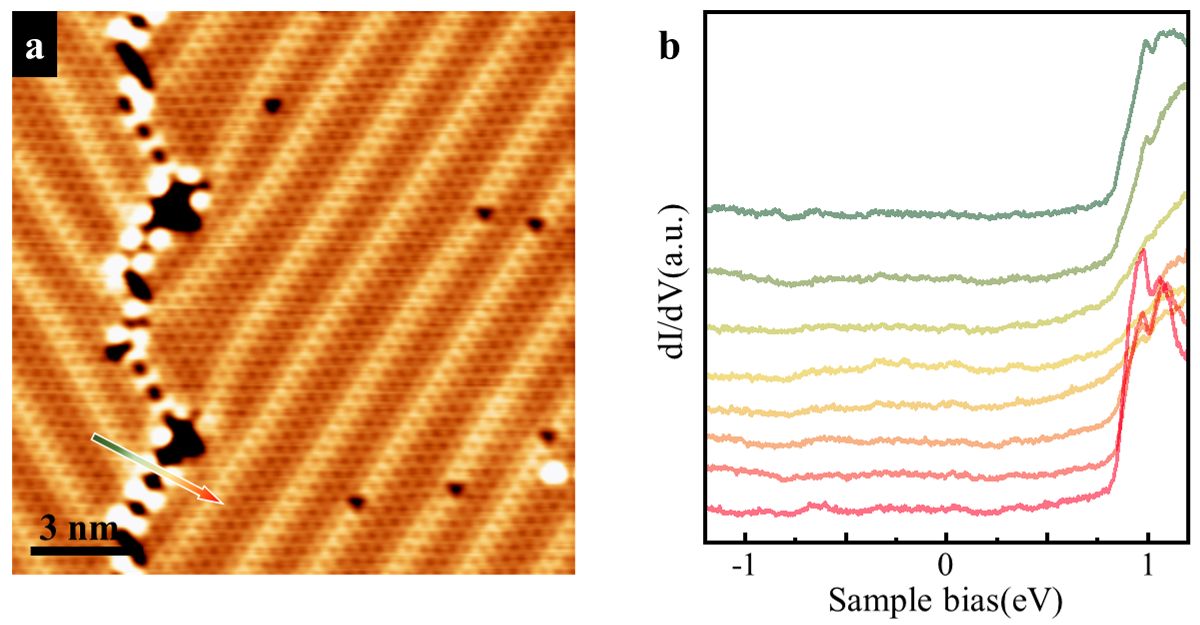}
    \caption{\textbf{The electronic properties crossing the line-defect formed by mirror-symmetric domains.} (a) A large-scale STM image of the line-defect. (b) The waterfall plot of the dI/dV curves along the gradient arrows in (a). Scanning parameters: (a) Vs = 0.4 V, It = 200 pA; (b) Vs = 2 V, It = 400 pA, Vrms = 10 mV.}
    \label{fig;S7}
\end{figure}

\begin{figure}[H]
\centering
\includegraphics[width=0.6\textwidth]{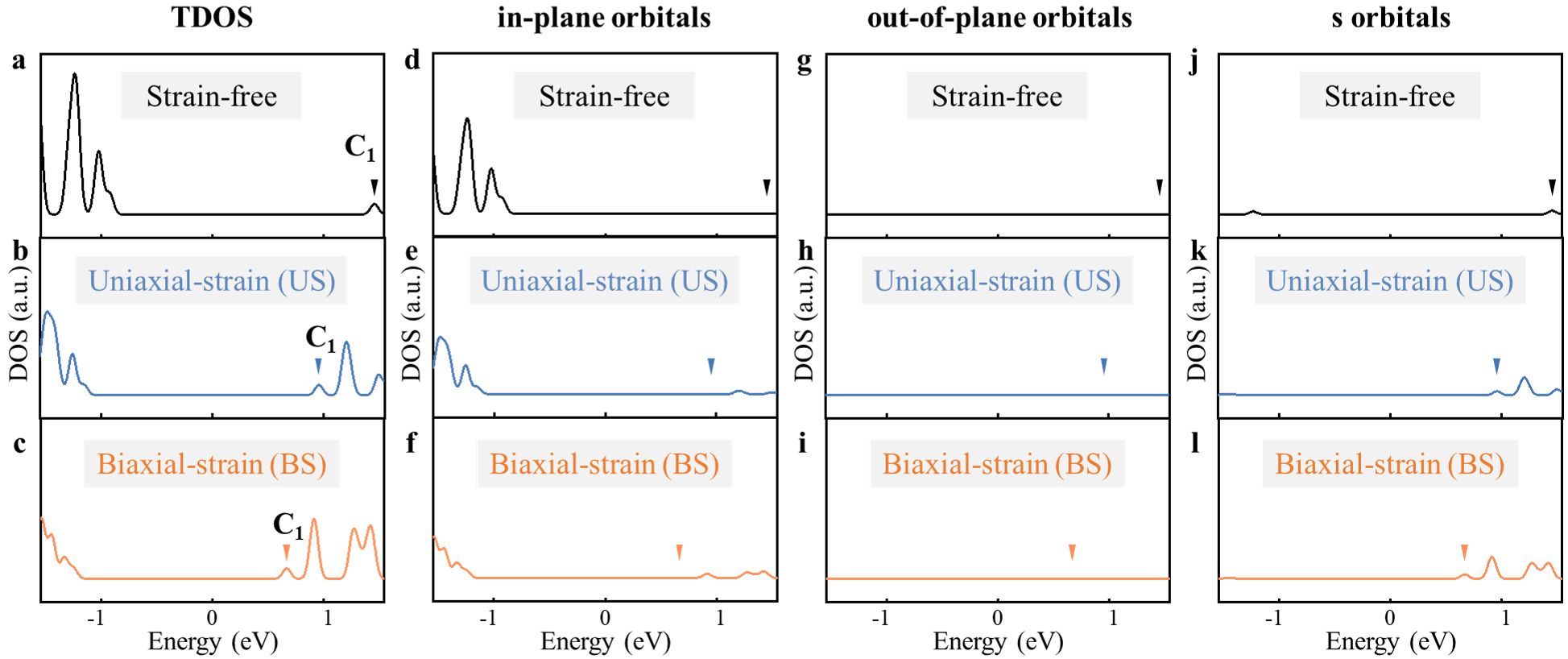}
    \caption{\textbf{Calculated PDOSs of monolayer CuSe under strain-free, 7.3\% uniaxial-strain and 7.3\% biaxial-strain on Cu(111). } (a-c) Total density of states (TDOSs). (d-f) PDOSs contributed by in-plane orbitals (Se $p_x/p_y$ and Cu $d_{xy}/d_{x^2-y^2}/d_{z^2}$). (g-i) PDOSs contributed by out-of-plane orbitals (Se $p_z$ and Cu $d_{xz}/d_{yz}$). (j-l) PDOSs contributed by s orbitals of Cu and Se atoms. }
    \label{fig;S8}
\end{figure}

\end{document}